\documentclass[conference]{IEEEtran}
\IEEEoverridecommandlockouts
\usepackage{cite}
\usepackage{amsmath,amssymb,amsfonts}
\usepackage{algorithmic}
\usepackage{graphicx}
\usepackage{textcomp}
\usepackage{xcolor}
\usepackage{caption}
\usepackage{subcaption}
\def\BibTeX{{\rm B\kern-.05em{\sc i\kern-.025em b}\kern-.08em
    T\kern-.1667em\lower.7ex\hbox{E}\kern-.125emX}}
\begin{document}

\title{Multimodal Breast Lesion Classification Using Cross-Attention Deep Networks
}

\makeatletter
\newcommand{\linebreakand}{%
  \end{@IEEEauthorhalign}
  \hfill\mbox{}\par
  \mbox{}\hfill\begin{@IEEEauthorhalign}
}
\makeatother

\author{\IEEEauthorblockN{Hung Q. Vo}
\IEEEauthorblockA{\textit{Electrical and Computer Engineering} \\
\textit{University of Houston}\\
Texas, USA \\
hqvo2@cougarnet.uh.edu}
\and
\IEEEauthorblockN{Pengyu Yuan}
\IEEEauthorblockA{\textit{Electrical and Computer Engineering} \\
\textit{University of Houston}\\
Texas, USA \\
pyuan2@central.uh.edu}
\and
\IEEEauthorblockN{Tiancheng He}
\IEEEauthorblockA{\textit{Systems Medicine and} \\
\textit{Biomedical Engineering} \\
\textit{Houston Methodist}\\
Texas, USA \\
THe@houstonmethodist.org}
\linebreakand
\IEEEauthorblockN{Stephen T.C. Wong}
 \IEEEauthorblockA{\textit{Systems Medicine and} \\
\textit{Biomedical Engineering} \\
\textit{Houston Methodist}\\
Texas, USA \\
STWong@houstonmethodist.org}
\and
\IEEEauthorblockN{Hien V. Nguyen}
\IEEEauthorblockA{\textit{Electrical and Computer Engineering} \\
\textit{University of Houston}\\
Texas, USA \\
hienvnguyen@uh.edu}
}

\maketitle

\begin{abstract}
Accurate breast lesion risk estimation can significantly reduce unnecessary biopsies and help doctors decide optimal treatment plans. Most existing computer-aided systems rely solely on mammogram features to classify breast lesions. While this approach is convenient, it does not fully exploit useful information in clinical reports to achieve the optimal performance. Would clinical features significantly improve breast lesion classification compared to using mammograms alone? How to handle missing clinical information caused by variation in medical practice? What is the best way to combine mammograms and clinical features? There is a compelling need for a systematic study to address these fundamental questions. This paper investigates several multimodal deep networks based on feature concatenation, cross-attention, and co-attention to combine mammograms and categorical clinical variables. We show that the proposed architectures significantly increase the lesion classification performance (average area under ROC curves from 0.89 to 0.94). We also evaluate the model when clinical variables are missing. 
\end{abstract}


\section{Introduction}
Breast cancer is responsible for over 42,000 women deaths in the United States and 685,000 deaths globally in 2020 \cite{Siegel2020}. Mammogram screening for early detection of breast lesions is important for decreasing the mortality rate \cite{broeders2012impact}. A major challenge in the screening procedure is the considerable inter-radiologist diagnostic performance variation \cite{mckinney2020international}. Both missing a cancer case or predicting a lesion to be malignant due to over-diagnosis would create severe consequences. False-negative cases could delay treatment and decrease the patient survival chance \cite{houssami2017epidemiology}. In contrast, false-positive diagnoses can lead to unnecessary biopsies which often create  stress, bleeding, bruising, and financial burden. 

The diagnostic performance of computer-aided software has significantly increased thanks to recent advances in deep learning. They hold great potential in helping radiologists to make more accurate diagnoses and reduce the performance variation \cite{freer2001screening}. Previous work focuses mostly on the general problem of classifying whether the lesion is benign or malignant \cite{shen2019deep, ribli2018detecting}. Recently, several papers on breast cancer have started to tackle more specific classification problems, including lesion types classification (mass or calcification) and BI-RADS density classification (level 1 to 4) \cite{shen2019deep, matthews2020multi, chougrad2020multi}. In \cite{shen2019deep}, given a lesion image, it will be categorized into five different classes: benign mass, malignant mass, benign calcification, malignant calcification or background. Masses are defined as three-dimensional space-occupying lesions in the breasts. Calcifications are deposits of calcium salts in the breast. Because different types of lesions (mass, calcification, etc.) have different properties, classifying the lesion type first can help the cancer diagnostic process. Breast density is another important factor for pathology evaluation \cite{kerlikowske2010breast}. In \cite{matthews2020multi}, they focused on this sub-problem and showed promising results. Chougrad et al. \cite{chougrad2020multi} combined the information of lesion types, breast density and pathology in a multi-label setting to exploit the correlation that may exist between those labels.



To the best of our knowledge, most approaches only use mammograms as input information for pathology classification. In this work, by using the available labels of clinical features that go along with their corresponding mammogram in the CBIS-DDSM database \cite{lee2017curated}, we propose a multimodal model that help improve the pathology classification performance by leveraging multimodal data. Further, for the problem of missing clinical features data which is often the case in real-world setting, we carry out evaluations for our proposed framework when it is combined with either co-attention or cross-attention.

Specifically, the contributions of this paper include:
\begin{enumerate}
    \item \textbf{Multimodal features combination for improving breast cancer pathology diagnostic}: we combine imaging features and clinical features for breast cancer diagnostic. Here, imaging features will be an embedding extracted from a pretrained deep learning model. For clinical features, we use a total of five features: breast density, mass shape, mass margins, calcification type, and calcification distribution. These features will be represented as a vector which is concatenated from 5 one-hot vectors (or a zero vector for each of missing feature), each vector is for one of the 5 features.
    \item \textbf{Evaluation of using co-attention and cross-attention in the presence of missing data}: in real-world setting, clinical features sometimes are either not provided or incomplete due to variations in clinical practice. To allow our model to be wisely adaptive to clinical situations so that it can be easily plugged to any clinical workflows, we incorporate either co-attention or cross-attention module to our proposed framework. We further train them to cope with missing clinical features.
\end{enumerate}

\section{Methodology}

\subsection{Multimodal Fusion Network Architecture}

\begin{figure}
    \centering
    \includegraphics[width=0.9\linewidth]{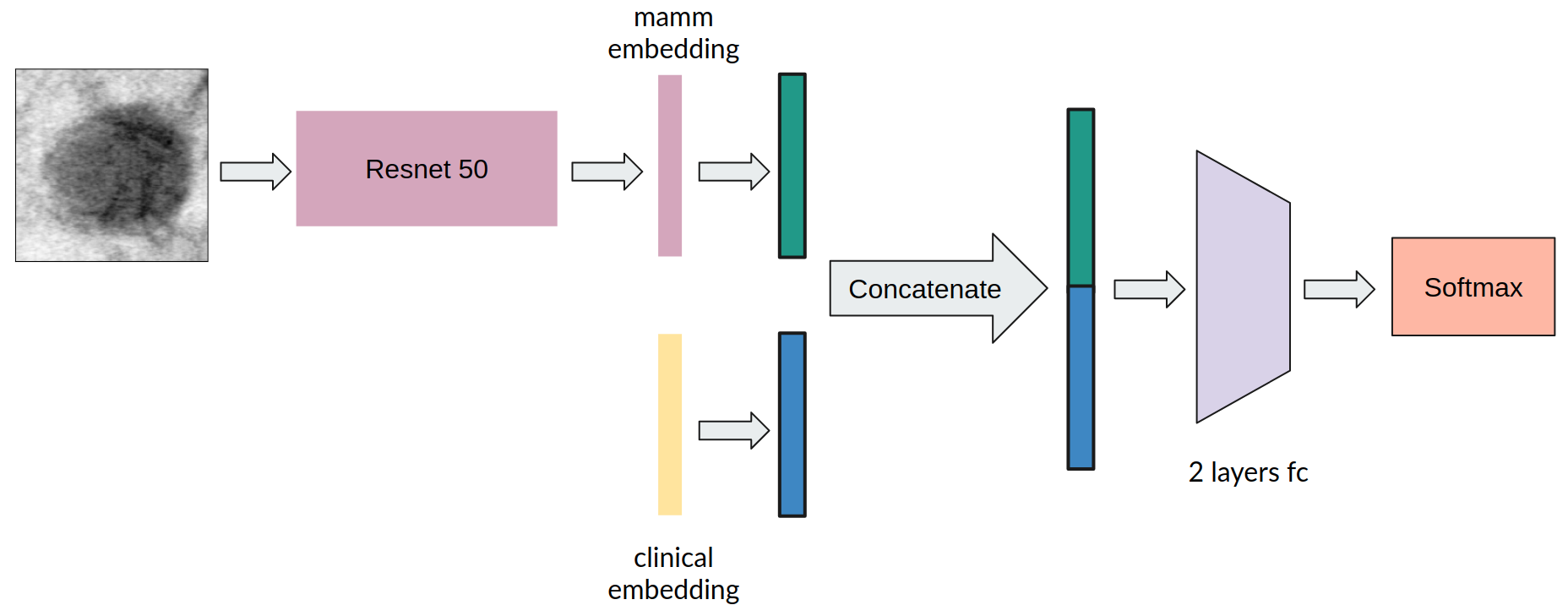}
    \caption{Overall multimodal fusion network architecture. Mammogram embedding is obtained by feeding a mammogram through a pretrained deep network. Both mammogram and clinical embedding are first projected into a fixed k-dimensional embedding. Then, two new projected embeddings are concatenated before feeding into a fully-connected network for classification.}
    \label{fig:framework}
\end{figure}

\subsubsection{Mammogram Embedding}
We use a ResNet-50 which has been pretrained on ImageNet to extract feature maps from mammograms. For each mammogram $x_i$, we obtain a 2048-d feature map $e_i$ from the global avarage pooling layer:

\begin{equation}
    e_i=ResNet50(x_i)
\end{equation}

In this work, we use ResNet50 as our model backbone but it can be replaced by any other architectures. The obtained feature map $e_i$ is furthed projected into a 100-dimensional embedding vector $\Tilde{e_i}$:

\begin{equation}
    \Tilde{e_i}=F(W_x^T\cdot{e_i}+b_x),
\end{equation}
where F is a nonlinear activation function, here ReLU is used.

\subsubsection{Clinical Embedding}
We use a total of five features for our clinical features: breast density, mass shape, mass margins, calcification type, and calcification distribution. Basically, mass shape and mass margins are only present for mass lesions. Similarly, calcification type and calcification distribution are only present for calcification lesions.
Each of the 5 features will be described by an one-hot vector. 
\begin{itemize}
    \item Breast Density - BI-RADS classifies breast density into four groups: entirely fatty, scattered fibroglandular densities, heterogeneously dense, and extremely dense. Thus, our one-hot vector will have four dimensions.
    \item Mass Shape - shape can receive value such as round, oval, irregular, and lobulated. Mass shape is basically categorized into eight classes so the representation will be an 8-dimensional one-hot vector.
    \item Mass Margins - margin can receive value such as ill defined, circumscribed, and spiculated. Mass margin is basically categorized into five classes so the representation will be a 5-dimensional one-hot vector.
    \item Calcification Type - type can receive value such as amorphous, punctate, and vascular. Calcification type is basically categorized into fourteen classes so the representation will be a 14-dimensional one-hot vector.
    \item Calcification Distribution - distribution can receive value such as clustered, linear, and regional. Calcification distribution is basically categorized into five classes so the representation will be a 5-dimensional one-hot vector.
\end{itemize}

In the cases of missing features, the corresponding vectors are set as zero vectors. For instance, mass lesions will neither have calcification type nor calcification distribution so their representations will be two zero vectors. This is similar for the case of calcification lesions. In real-world applications, any missing feature will result in one zero vector. In summary, our clinical embedding will have 36 dimensions, which is the sum of dimensions of 5 feature vectors.

Each clinical embedding $c_i$ is projected into a 100-dimensional embedding $\Tilde{c_i}$, which is the same as for mammogram embedding:

\begin{equation}
    \Tilde{c_i}=F(W_t^T\cdot{c_i}+b_t)
\end{equation}

\subsubsection{Multimodal fusion}
After projected mammogram embedding $\Tilde{e_i}$ and projected clinical embedding $\Tilde{c_i}$ are obtained, they will be concatenated before feeding into 2 fully connected layers for pathology classification:

\begin{equation}
    k_i=Concat(\Tilde{e_i}, \Tilde{c_i}),
\end{equation}
where $k_i$ is the final concatenated embedding. There are multiple ways to fuse information, e.g., early fusion and late fusion. For simplicity, in this work, we simply concatenate two projected embedding vectors but other methods could be experimented to see whether they further improve the model performance. Figure \ref{fig:framework} shows the overall framework of our proposed multimodal fusion model.

\subsection{Co-attention and Cross-attention Modules}

\begin{figure}
    \centering
    \includegraphics[width=0.9\linewidth]{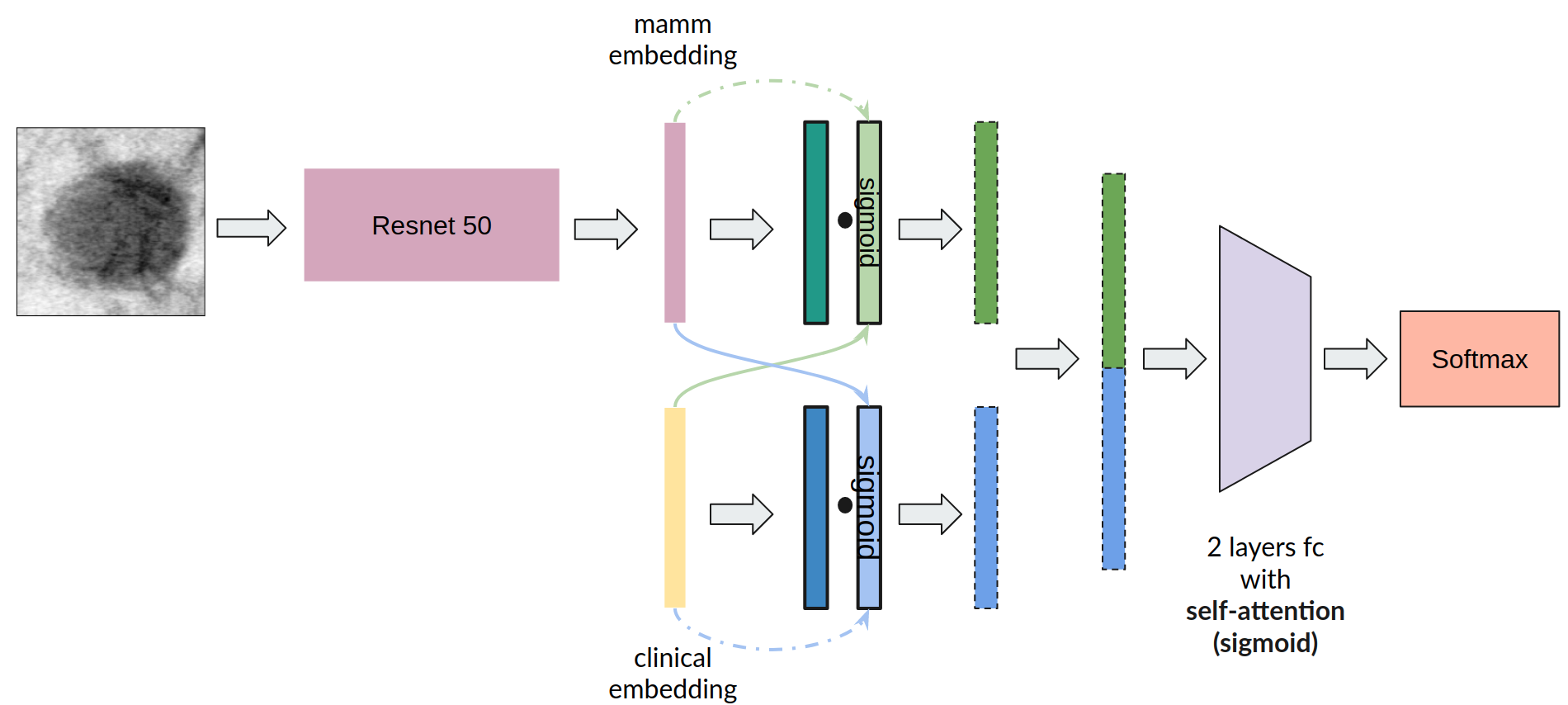}
    \caption{Network architecture with co-attention/cross-attention. We incorporate either co-attention or cross-attention module to our proposed framework. For cross-attention module, two dashed arrows in the figure are removed.}
    \label{fig:att_framework}
\end{figure}

In clinical settings, mammograms are not often provided together with clinical features or only part of the features are provided. This leads to the problems of missing values. In this work, we make evaluations of our model when it is combined with either co-attention or cross-attention module. Here, co-attention/cross-attention is used with the intention of teaching the model to learn the relevant information between a mammogram and its clinical features. Figure \ref{fig:att_framework} shows our proposed framework when combined with co-attention/cross-attention module.

\subsubsection{Co-attention Module}
Co-attention was proposed in \cite{lu2019vilbert}. In this paper, being inspired by the idea of co-attention, we try to combine the projected mammogram embedding $\Tilde{e_i}$ and the projected clinical embedding $\Tilde{c_i}$ as follows:
\begin{equation}
\begin{split}
    \alpha_{e_i}=\sigma(W_x^{'T}\cdot{Concat(e_i, c_i)} + b_v^{'}),\\
    \alpha_{c_i}=\sigma(W_t^{'T}\cdot{Concat(e_i, c_i)} + b_t^{'}),\\
    e_i^{aug}=\alpha_{e_i}\odot\Tilde{e_i}, \quad c_i^{aug}=\alpha_{c_i}\odot\Tilde{c_i},\\
    k_i=Concat(e_i^{aug}, c_i^{aug}),
\end{split}
\end{equation}
where $\sigma$ is the sigmoid activation function, $\alpha_{e_i}$ and $\alpha_{c_i}$ are the augmented coefficients, $k_i$ is the final concatenated embedding.

\subsubsection{Cross-attention Module}
Cross-attention was proposed in \cite{abavisani2020multimodal}. Basically, the only difference between co-attention and cross-attention is in the way of calculating the augmented coefficients. While co-attention uses information from both the multimodal data, cross-attention only uses the information in the other modality to compute the coefficient for the current modality, which is as follow:

\begin{equation}
\begin{split}
    \alpha_{e_i}=\sigma(W_x^{'T}\cdot{c_i} + b_v^{'}), \quad
    \alpha_{c_i}=\sigma(W_t^{'T}\cdot{e_i} + b_t^{'})
\end{split}
\end{equation}

\subsubsection{Missing Clinical Variables}
Hospitals have different practices, therefore, might not use the same set of clinical variables. Even for the same hospital, patient information often has missing values. Therefore, it is important to study the effect of missing variables on our model's performance. We use bait-and-switch training where clinical features are removed from randomly selected mammograms during both training and testing phases. 





\begin{figure*}
    \centering
    \begin{subfigure}[c]{0.37\textwidth}
        \resizebox{\linewidth}{!}{
        \setlength{\tabcolsep}{5pt}
        \begin{tabular}{cc}
        
        \shortstack{\textbf{{\scriptsize Mammogram Only}} \\ \includegraphics[width=0.7\linewidth]{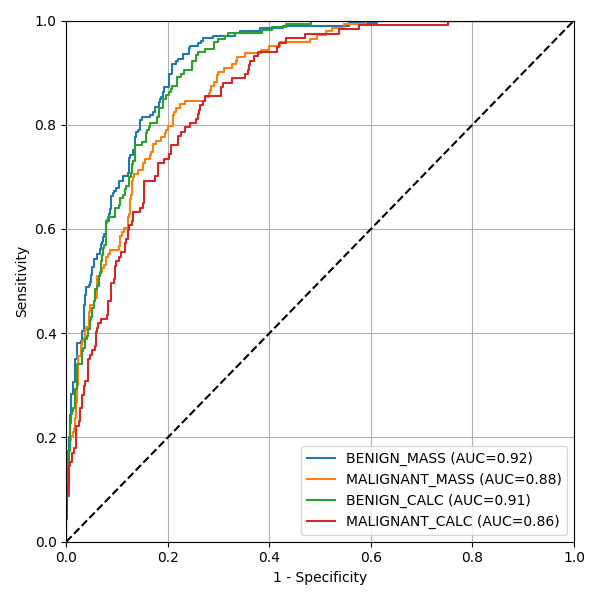}}&
        \shortstack{\textbf{{\scriptsize With Clinical Features}} \\ \includegraphics[width=0.7\linewidth]{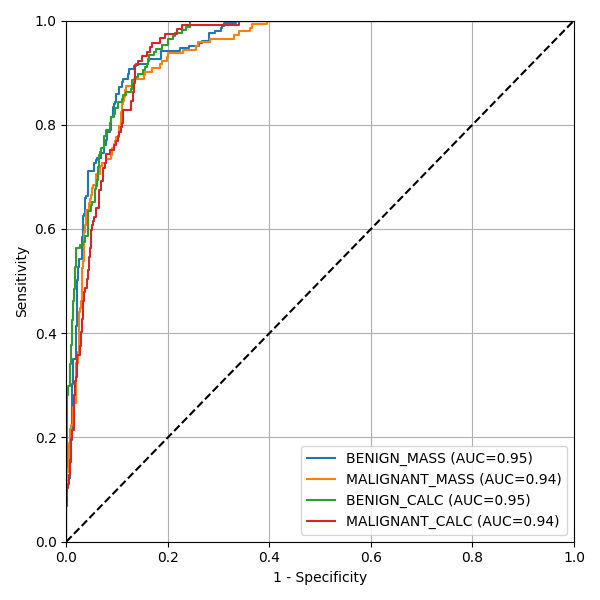}} \\
        
        \shortstack{ 
        \includegraphics[width=0.7\linewidth]{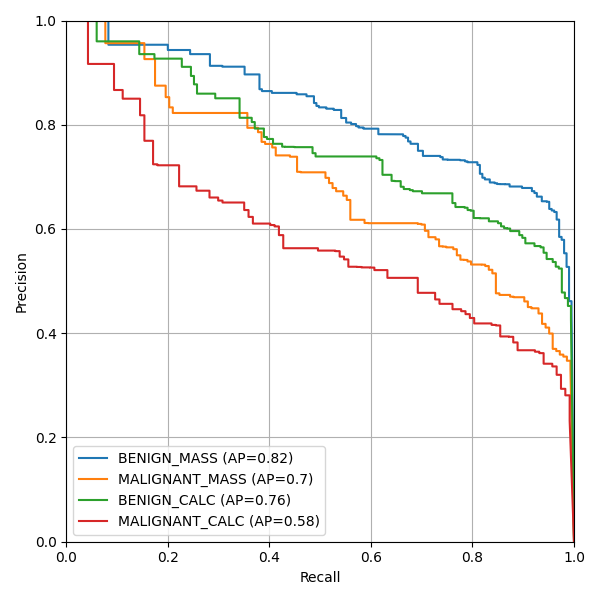}}&
        \shortstack{
        \includegraphics[width=0.7\linewidth]{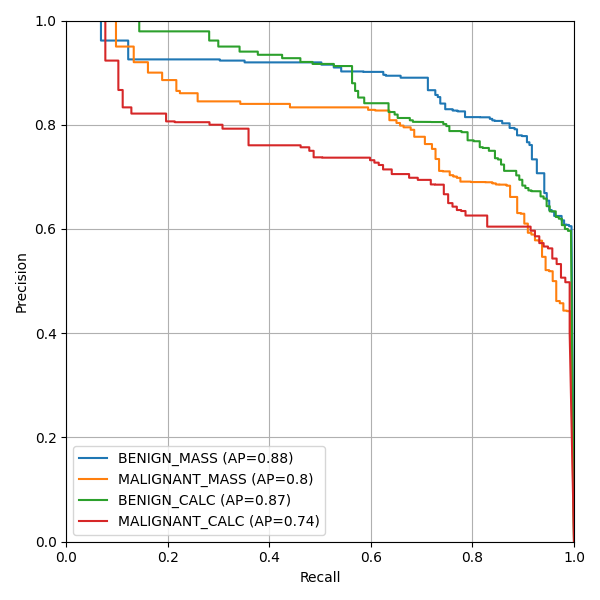}} \\
        
        \end{tabular}
        }
        
        \caption{Results comparison when using additional clinical features.}
        \label{fig:origin_vs_with_clinical_feats}
    \end{subfigure}
    \hfill
    \begin{subfigure}[c]{0.52\textwidth}
        \centering
        \resizebox{\linewidth}{!}{
        \setlength{\tabcolsep}{7pt}
        \begin{tabular}{ccc}
        
        \shortstack{\textbf{Concat} \\ \includegraphics[width=0.7\linewidth]{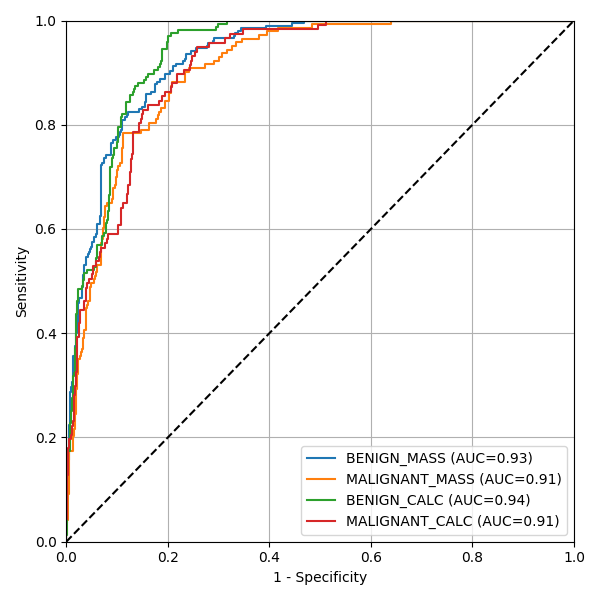}}&
        \shortstack{\textbf{Co-attention} \\ \includegraphics[width=0.7\linewidth]{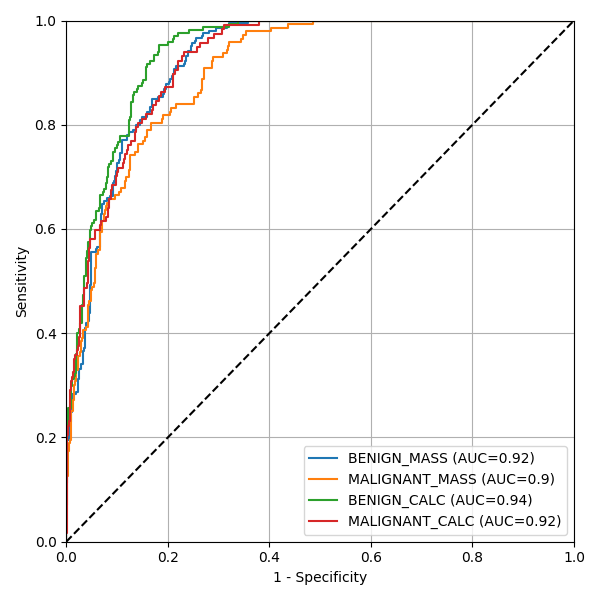}}&
        \shortstack{\textbf{Cross-attention} \\ \includegraphics[width=0.7\linewidth]{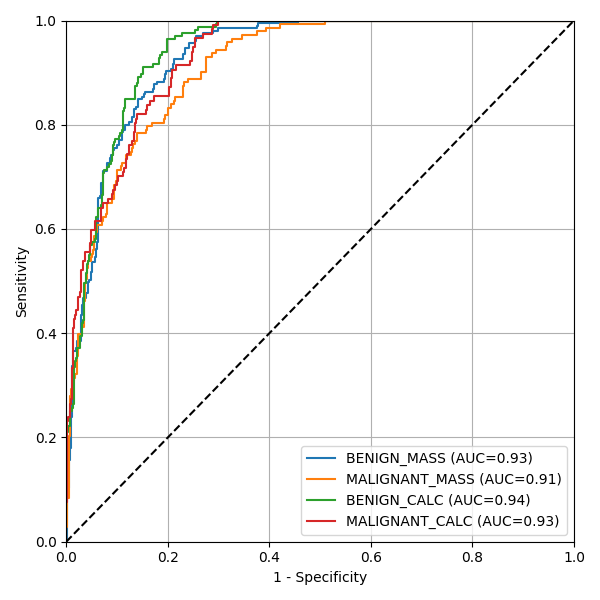}} \\
        
        \shortstack{ 
        \includegraphics[width=0.7\linewidth]{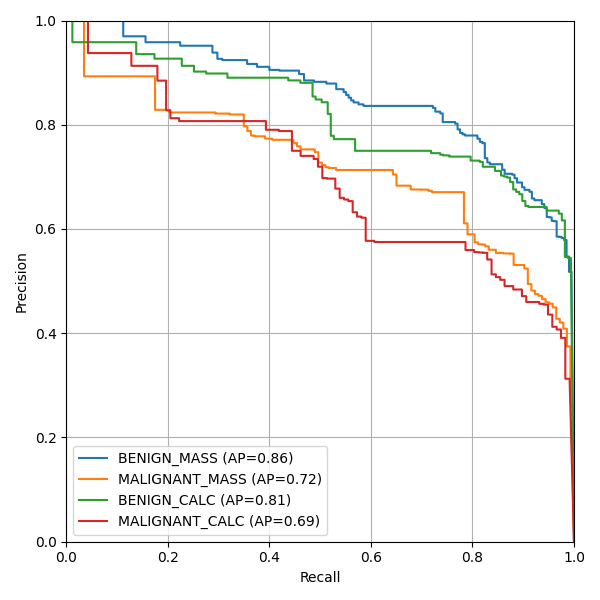}}&
        \shortstack{ 
        \includegraphics[width=0.7\linewidth]{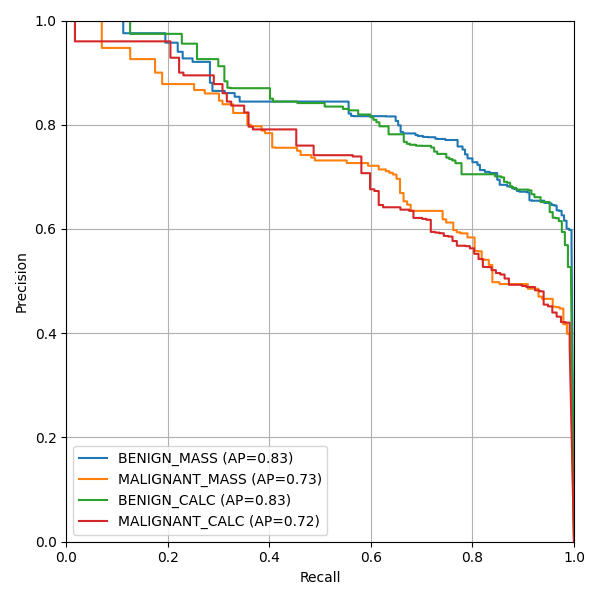}}&
        \shortstack{
        \includegraphics[width=0.7\linewidth]{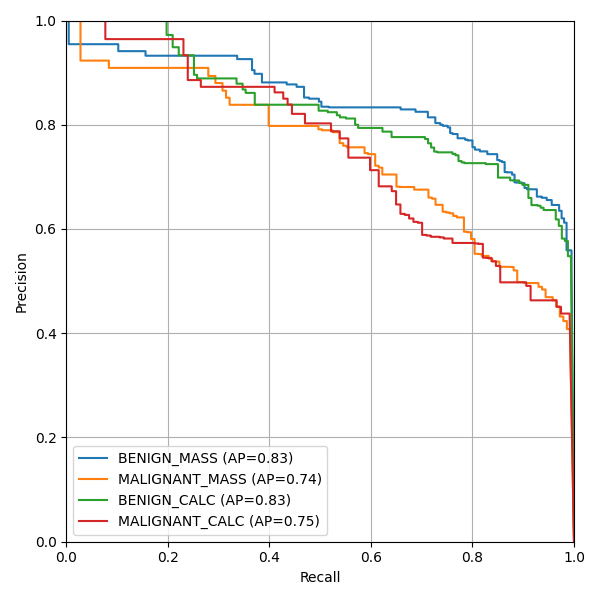}} \\
        
        \end{tabular}
        }
        
        \caption{Bait-and-Switch Training: Here, the probability of mammograms
        whose clinical data are removed is set to be 0.3 for both training and testing.}
        \label{fig:bait_and_switch}
    \end{subfigure}
    
    \caption{ROC curves and PR curves evaluated on the official CBIS-DDSM test set.}
    \label{fig:models_evaluation}
    \vspace{-3mm}
\end{figure*}

\begin{figure}
    \centering
    \includegraphics[width=0.9\linewidth]{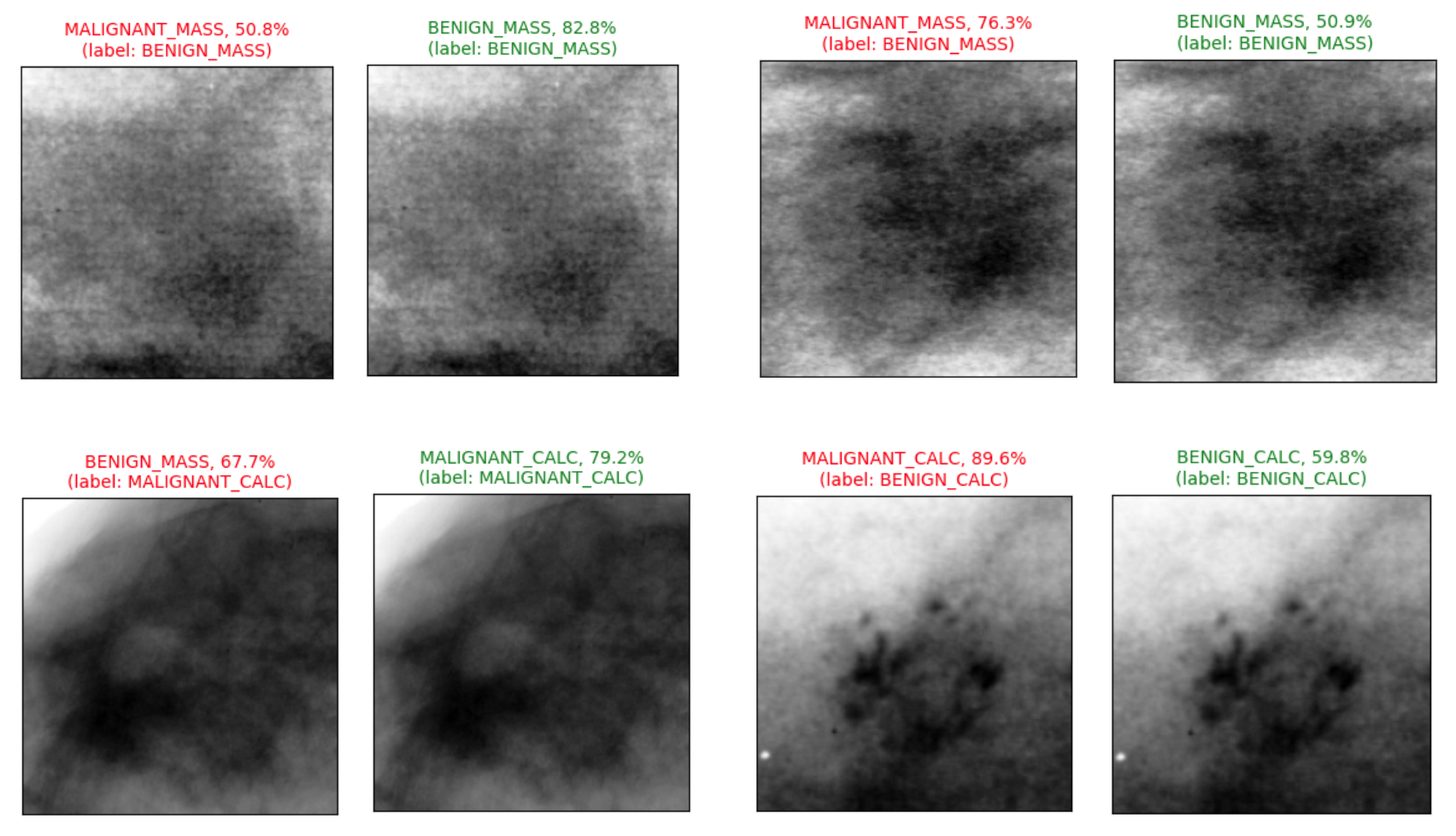}
    \caption{Visualization of four cases when the original model fails to classify but using additional clinical features can help. For each case, on the left is the prediction of the original model while on the right is for the model that uses additional clinical features.}
    \label{fig:visualize_fail_cases}
\end{figure}


\section{Experimental Results}
\subsection{Breast Lesion Dataset}
Our experiments use the CBIS-DDSM dataset that consists of 1592 mass mammograms and 1511 calcification mammograms. The official data train test split is 1231/361 for the mass cases and 1227/284 for the calcification cases. We further split the official training set into 75\% for training and 25\% for validation. We only consider clinical features with single label because the number of multi-label lesions is very small. We use the bounding box inferred from mask ground-truth to crop the lesion from mammogram without any further pre-processing steps.

\subsection{Hyperparamter Settings}
We trained our network using the setting in \cite{shen2019deep}. We used ResNet50 as our backbone and we finetuned the network for three stages with more layers are gradually unfrozen through each stage. We used Adam Optimizer and the learning rates are set to 1e-3, 1e-4, 1e-5 for 1$^{st}$ stage, 2$^{nd}$ stage and 3$^{rd}$ stage. We used the same setting for the multimodal networks. For these networks, both the projected mammogram embedding and projected clinical embedding will have 100 dimensions. For the two fully-connected layers, we set them to have an equal of 200 neurons. Our training used the same data augmentation procedure in \cite{shen2019deep}.

\subsection{Multimodal Classification}
The result when using additional clinical features is shown in Fig.~\ref{fig:origin_vs_with_clinical_feats}. Both the areas under ROC curves (AUC-ROC) and Precision-Recall curves (AUC-PR) significantly increase when clinical features are used. In particular, the average AUC-ROC increases from 0.892 to 0.945, and the average AUC-PR from 0.715 to 0.82 when using attention deep networks. Our results indicate that using additional clinical features can help boost up the pathology classification performance to a large margin. Figure \ref{fig:visualize_fail_cases} shows four example cases when the prediction is incorrect if only mammogram is used, but can be correctly predicted if clinical features are used. Some of the cases which have been misclassified as benign, can be now recognized as malignant by leveraging the clinical information. Three fusion approaches perform competitively well, when co-attention and cross-attention have a slight edge over the traditional feature concatenation method.

\subsection{Effects of Missing Variables}
 In our experiments, we set the probability of mammograms whose clinical data are removed to be 0.3 for both training and testing. We evaluated on three different fusion approaches: normal concatenation, with co-attention, and with cross-attention. The results are shown in Figure \ref{fig:bait_and_switch}. For the ROC curves, it seems there are no significant differences among three fusion settings. Nonetheless, the PR curves have shown promising results for using attention to deal with missing data. The AUC-PR of malignant mass and malignant calcification, which are 0.72 and 0.69 for normal concatenation setting, increase up to 0.73 and 0.72 for co-attention setting. They are further increased when using cross-attention, to the extent of 0.74 and 0.75, respectively. It is worth mentioning that for benign mass, its AUC-PR has slightly decreased when attention modules are used (from 0.86 to 0.83).

\section{Conclusion}
This paper studies multimodal deep networks based on simple concatenation, cross-attention, and co-attention to combine mammography and clinical features. We demonstrate that our multimodal approach significantly increases the breast lesion classifier's area under curves and thus the diagnostic precision. We also examine the model's performance under the scenario of missing variables due to variations in clinical practice. Our future work will explore the feasibility of extracting the selected clinical features directly from mammograms. \vspace{5mm}

\textbf{Acknowledgement:} This research is funded by NIH R01CA251710, T.T. and W.F. Chao Foundation and John S
Dunn Research Foundation.

\bibliographystyle{IEEEtran}
\bibliography{IEEEabrv,ref}

\begin{thebibliography}{10}
\providecommand{\url}[1]{#1}
\csname url@samestyle\endcsname
\providecommand{\newblock}{\relax}
\providecommand{\bibinfo}[2]{#2}
\providecommand{\BIBentrySTDinterwordspacing}{\spaceskip=0pt\relax}
\providecommand{\BIBentryALTinterwordstretchfactor}{4}
\providecommand{\BIBentryALTinterwordspacing}{\spaceskip=\fontdimen2\font plus
\BIBentryALTinterwordstretchfactor\fontdimen3\font minus
  \fontdimen4\font\relax}
\providecommand{\BIBforeignlanguage}[2]{{%
\expandafter\ifx\csname l@#1\endcsname\relax
\typeout{** WARNING: IEEEtran.bst: No hyphenation pattern has been}%
\typeout{** loaded for the language `#1'. Using the pattern for}%
\typeout{** the default language instead.}%
\else
\language=\csname l@#1\endcsname
\fi
#2}}
\providecommand{\BIBdecl}{\relax}
\BIBdecl

\bibitem{Siegel2020}
\BIBentryALTinterwordspacing
R.~L. Siegel, K.~D. Miller, and A.~Jemal, ``Cancer statistics, 2020,''
  \emph{CA: A Cancer Journal for Clinicians}, vol.~70, no.~1, pp. 7--30, 2020.
  [Online]. Available:
  \url{https://acsjournals.onlinelibrary.wiley.com/doi/abs/10.3322/caac.21590}
\BIBentrySTDinterwordspacing

\bibitem{broeders2012impact}
M.~Broeders, S.~Moss, L.~Nystr{\"o}m, S.~Njor, H.~Jonsson, E.~Paap, N.~Massat,
  S.~Duffy, E.~Lynge, and E.~Paci, ``The impact of mammographic screening on
  breast cancer mortality in europe: a review of observational studies,''
  \emph{Journal of medical screening}, vol.~19, no. 1\_suppl, pp. 14--25, 2012.

\bibitem{mckinney2020international}
S.~M. McKinney, M.~Sieniek, V.~Godbole, J.~Godwin, N.~Antropova, H.~Ashrafian,
  T.~Back, M.~Chesus, G.~S. Corrado, A.~Darzi \emph{et~al.}, ``International
  evaluation of an ai system for breast cancer screening,'' \emph{Nature}, vol.
  577, no. 7788, pp. 89--94, 2020.

\bibitem{houssami2017epidemiology}
N.~Houssami and K.~Hunter, ``The epidemiology, radiology and biological
  characteristics of interval breast cancers in population mammography
  screening,'' \emph{NPJ Breast Cancer}, vol.~3, no.~1, pp. 1--13, 2017.

\bibitem{freer2001screening}
T.~W. Freer and M.~J. Ulissey, ``Screening mammography with computer-aided
  detection: prospective study of 12,860 patients in a community breast
  center,'' \emph{Radiology}, vol. 220, no.~3, pp. 781--786, 2001.

\bibitem{shen2019deep}
L.~Shen, L.~R. Margolies, J.~H. Rothstein, E.~Fluder, R.~McBride, and W.~Sieh,
  ``Deep learning to improve breast cancer detection on screening
  mammography,'' \emph{Scientific reports}, vol.~9, no.~1, pp. 1--12, 2019.

\bibitem{ribli2018detecting}
D.~Ribli, A.~Horv{\'a}th, Z.~Unger, P.~Pollner, and I.~Csabai, ``Detecting and
  classifying lesions in mammograms with deep learning,'' \emph{Scientific
  reports}, vol.~8, no.~1, pp. 1--7, 2018.

\bibitem{matthews2020multi}
T.~P. Matthews, S.~Singh, B.~Mombourquette, J.~Su, M.~P. Shah, S.~Pedemonte,
  A.~Long, D.~Maffit, J.~Gurney, R.~Morales~Hoil \emph{et~al.}, ``A multi-site
  study of a breast density deep learning model for full-field digital
  mammography images and synthetic mammography images,'' \emph{Radiology:
  Artificial Intelligence}, p. e200015, 2020.

\bibitem{chougrad2020multi}
H.~Chougrad, H.~Zouaki, and O.~Alheyane, ``Multi-label transfer learning for
  the early diagnosis of breast cancer,'' \emph{Neurocomputing}, vol. 392, pp.
  168--180, 2020.

\bibitem{kerlikowske2010breast}
K.~Kerlikowske, A.~J. Cook, D.~S. Buist, S.~R. Cummings, C.~Vachon, P.~Vacek,
  and D.~L. Miglioretti, ``Breast cancer risk by breast density, menopause, and
  postmenopausal hormone therapy use,'' \emph{Journal of Clinical Oncology},
  vol.~28, no.~24, p. 3830, 2010.

\bibitem{lee2017curated}
R.~S. Lee, F.~Gimenez, A.~Hoogi, K.~K. Miyake, M.~Gorovoy, and D.~L. Rubin, ``A
  curated mammography data set for use in computer-aided detection and
  diagnosis research,'' \emph{Scientific data}, vol.~4, no.~1, pp. 1--9, 2017.

\bibitem{lu2019vilbert}
J.~Lu, D.~Batra, D.~Parikh, and S.~Lee, ``Vilbert: Pretraining task-agnostic
  visiolinguistic representations for vision-and-language tasks,'' \emph{arXiv
  preprint arXiv:1908.02265}, 2019.

\bibitem{abavisani2020multimodal}
M.~Abavisani, L.~Wu, S.~Hu, J.~Tetreault, and A.~Jaimes, ``Multimodal
  categorization of crisis events in social media,'' in \emph{Proceedings of
  the IEEE/CVF Conference on Computer Vision and Pattern Recognition}, 2020,
  pp. 14\,679--14\,689.

\end{thebibliography}

\end{document}